\documentclass[10pt]{bmc_article}    
\usepackage{cite}
\usepackage{url}
\usepackage{ifthen}  % Conditional 
\usepackage{booktabs,multicol,multirow}
\usepackage{array}
\usepackage{epsfig}
\usepackage{amsmath,amssymb,amsfonts,amscd}
\usepackage[bf]{caption}
\usepackage{bm}% bold math

\newcommand{\be}{\begin{equation}}
\newcommand{\ee}{\end{equation}}
\newcommand{\bea}{\begin{eqnarray}}
\newcommand{\eea}{\end{eqnarray}}
\newboolean{publ}

\newenvironment{bmcformat}{\fussy\setboolean{publ}{true}}{\fussy}

\begin{document}
\begin{bmcformat}

\title{Genetic noise control via protein oligomerization}

\author{
Cheol-Min Ghim\email{C.-M. Ghim - cmghim@llnl.gov} 
and Eivind Almaas\email{E. Almaas - almaas@llnl.gov}\correspondingauthor
}

\address{Microbial Systems Biology Group,
Biosciences and Biotechnology Division,
Lawrence Livermore National Laboratory, 
7000 East Avenue Livermore, CA 94550, USA}

\maketitle
\begin{abstract}
\paragraph*{Background:} 
Gene expression in a cell entails random reaction events occurring
over disparate time scales.  Thus, molecular noise that often results
in phenotypic and population-dynamic consequences sets a fundamental
limit to biochemical signaling.  While there have been numerous
studies correlating the architecture of cellular reaction networks
with noise tolerance, only a limited effort has been made to
understand the dynamic role of protein-protein interactions.  
\paragraph*{Results:} We have developed a fully stochastic 
model for the positive feedback control of a single gene, as well as a
pair of genes (toggle switch), integrating quantitative results from
previous \textit{in vivo} and \textit{in vitro} studies.  In
particular, we explicitly account for the fast binding-unbinding kinetics 
among proteins, RNA polymerases, and the promoter/operator sequences of DNA.  
We find that the overall noise-level is reduced and the frequency content 
of the noise is dramatically shifted to the physiologically irrelevant 
high-frequency regime in the presence of protein dimerization.  
This is independent of the choice of monomer or dimer as transcription 
factor and persists throughout the multiple model topologies considered.  
For the toggle switch, we additionally find that the presence of a protein 
dimer, either homodimer or heterodimer, may significantly reduce its 
random switching rate.
Hence, the dimer promotes the robust function of bistable switches by
preventing the uninduced (induced) state from randomly being induced
(uninduced).
\paragraph*{Conclusions:} The specific binding between regulatory proteins
provides a buffer that may prevent the propagation of fluctuations in
genetic activity.  The capacity of the buffer is a non-monotonic
function of association-dissociation rates.  Since the protein
oligomerization \textit{per se} does not require extra protein
components to be expressed, it provides a basis for the rapid control
of intrinsic or extrinsic noise.  The stabilization of regulatory 
circuits and epigenetic memory in general is of direct implications to 
organism fitness. Our results also suggest possible avenues for the design 
of synthetic gene circuits with tunable robustness for a wide range of 
engineering purposes.
\end{abstract}

\ifthenelse{\boolean{publ}}{\begin{multicols}{2}}{}
\section*{Background} 
Recent experiments on isogenic populations of microbes with
single-cell resolution~\cite{pedraza05,rosenfeld05,newman06} have
demonstrated that stochastic fluctuations, or noise, can override
genetic and environmental determinism.  In fact, the presence of noise
may significantly affect the fitness of an organism~\cite{raser05}.
The traditional approach for modeling the process of molecular
synthesis and degradation inside a cell is by deterministic rate
equations, where the continuous change of arbitrarily small fractions
of molecules is controlled instantaneously and frequently represented
through sigmoidal dose-response relations.  However, the rate-equation
approaches can not explain the observed phenotypic variability in an
isogenic population in stable environments.  In particular, when
molecules involved in feedback control exist in low copy numbers,
noise may give rise to significant cell-to-cell variation as many
regulatory events are triggered by molecules with very low copy
numbers $\lesssim100$~\cite{guptasarma95}.  A well known example is
the regulation of inorganic trace elements~\cite{nelson99}, such as
iron, copper, and zinc.  While these trace elements are essential for
the activity of multiple enzymes, their presence may quickly turn
cytotoxic unless their concentrations are carefully controlled.

Although the presence of phenotypic variation due to stochastic
fluctuations need not be detrimental for a population of
cells~\cite{kussell05}, elaborate regulatory mechanisms have evolved
to attenuate noise~\cite{chen06}.
Several systems-biology studies have
recently focused on a select set gene-regulatory circuits, in
particular those with feedback control.  Feedback control circuits
have been identified as important for multiple species and proven
responsible for noise reduction and increased functional stability in
many housekeeping genes through negative
autoregulation~\cite{thieffry98}, long cascades of ultrasensitive
signaling~\cite{thattai02}, bacterial chemotaxis~\cite{yi00}, and the
circadian clock~\cite{vilar02}.  Additionally, recent studies on iron
homeostasis~\cite{semsey06,levine07} in \textit{E. coli} highlight the
noise-reducing capability mediated by small RNAs.

Here, we study reversible protein-protein binding as a novel 
source for genetic noise control. In particular, we have quantitatively 
analyzed the effects of protein oligomerization on noise in positive 
autoregulatory circuits as well as a simple toggle-switch~\cite{gardner00}.  
The all-or-none threshold behavior of positive-feedback circuits typically 
improves robustness against ``leaky'' switching. However, due to their 
functional purposes, gene circuits involved in developmental processes 
or stress responses that often accompany genome-wide changes in gene 
expression are intrinsically noisier than negative feedback circuits.

It is frequently observed that transcription factors exist in
oligomeric form~\cite{beckett01}, and protein oligomerization is an
important subset of protein-protein interactions, constituting a
recurring theme in enzymatic proteins as well as regulatory proteins.
Well studied examples include the $\lambda$-phage repressor,
${\lambda}$CI (dimer), the TrpR (dimer), LacR (tetramer), and Lrp
(hexadecamer or octamer).  While many of the RNA-binding proteins
dimerize exclusively in the cytosol, the LexA repressor~\cite{kim92},
the leucine-zipper activator~\cite{berger98,kohler99}, and the Arc
repressor~\cite{rentzeperis99} have been shown to form an oligomer
either in the cytosol (``dimer path'') or on the DNA by sequential
binding (``monomer path'').  Previously, the efficacy of monomer 
and dimer transcription-regulation paths to reduce noise was separately
studied for a negative-feedback autoregulatory
circuit~\cite{bundschuh03jtb}.  In contrast, we have focused on
oligomerization in positive-feedback autoregulatory circuits, as well
as genetic toggle switches based on the mutual repression of
genes\cite{gardner00}. We find that cytosolic transcription-factor
oligomerization acts as a significant buffer for
abundance-fluctuations in the monomer, overall reducing noise in the
circuit.  Additionally, the noise-power spectral density is shifted
from the low- to the high-frequency regime.  In the toggle switch,
cytosolic oligomerization may significantly stabilize the functional
state of the circuit.  This is especially evident for heterodimerization.

Yet another interesting case of ligand-binding-mediated receptor
oligomerization has been reported~\cite{alarcon06,macnamara07}, where
the formation of various structures of oligomers may act to buffer the
intracellular signaling against noise. Although our modeling and
analysis is based on prokaryotic cells, we expect our main findings to
be organism independent since protein oligomers, especially
homodimers, is such a common occurrence across the
species~\cite{ispolatov05}, with homodimers comprising 12.6\% of the
high-fidelity human proteome~\cite{ramirez07,mcdermott05}.

\section*{Results and Discussion}
\subsection*{Dimerization breaks long-time noise correlations in 
autogenous circuit}
To evaluate the dynamic effects of protein-protein binding in
positive-autoregulation gene circuits, we construct several
alternative models of positive autogenous circuits.  Each model
emphasizes a different combination of possible feedback mechanisms,
and the network topologies considered can be grouped into the two
classes of monomer-only (MO) and dimer-allowed (DA) circuits,
according to the availability of a protein-dimer state (color coding
in Fig.~1).  We further group the DA circuits into three
variations, DA1 through DA3, depending on which form of the protein is
the functional transcription factor (TF) and where the dimerization
occurs.  For DA1, we only allow the dimer to bind with the DNA-operator
sequence (dimeric transcription factor, DTF), while for DA2
dimerization occurs through sequential binding of monomers on the DNA.
In DA3, the protein-DNA binding kinetics is the same as in the MO
circuit, hence monomeric transcription factor (MTF), with the addition
of a cytosolic protein dimer state.  While we will only present
results for DA1 in this paper, there is no significant difference for
DA2 and DA3 [Additional file 1].

Note that the feedback loop is not explicit in Fig.~1 but
implicitly included through the dependence of RNAp-promoter binding
equilibrium on the binding status of the TF-operator pair. The sign
(positive or negative) and strength of the feedback control is
determined by the relative magnitude of the dissociation constants
between RNAp and DNA which is either free or TF-bound.  For instance,
topology DA1 has positive feedback control if $K_{30}= k_{30}/q_{30} >
K_{32} = k_{32}/q_{32}$, and $K_{30}$ corresponds to the level of
constitutive transcription (transcription initiation in the absence of
bound transcription factor).  For each topology, we study the
dependence of noise characteristics on the kinetic rates by varying
the dimer lifetime, binding affinity, and the individual
association/dissociation rates (see Table~\ref{rates} and
Fig.~1).  While we only discuss positive feedback control
of the autogenous circuit in this paper, we have obtained
corresponding results for negative feedback control [Additional file 1].

Fig.~2 shows a sample of ten representative time courses
for the protein abundance.  The effect of stochastic fluctuations is
marked in the MO circuit.  However, in all the DA circuits where the
protein may form a cytosolic dimer we observe a significantly reduced
level of noise in the monomer abundance.  The suppression of
fluctuations persists throughout the range of kinetic parameters that
(so far) is known to be physiologically relevant (see
Table~1).

Calculating the steady-state distribution for the monomer and dimer
abundances (Fig.~3) we observe a clear trend that the
monomer Fano factor (variance-to-mean ratio) is reduced as the binding
equilibrium is shifted towards the dimer.  This trend is conserved for
all the investigated DA topologies (see \textit{Supplementary
Information}).  As long as dimerization is allowed in the cytosol, the
fast-binding equilibrium absorbs long-time fluctuations stemming from
bursty synthesis or decay of the monomer.  When a random fluctuation
brings about a sudden change in the monomer copy number, dimerization
provides a buffering pool that absorbs the sudden change.  Otherwise,
random bursts in the monomer abundance will propagate to the
transcriptional activity of the promoter, leading to erratic control
of protein expression.  It should be emphasized that this has nothing
to do with the sign of regulation and is in agreement with the
observations of Ref.~\cite{bundschuh03jtb} for negative
autoregulation.  Surprisingly, the magnitude of noise reduction in the
positive autoregulatory circuit is nearly the same as that for
negative autoregulation which is typically considered a highly stable
construct [Additional file 1].

A heuristic explanation can be found from Jacobian analysis of a
deterministic dynamical system, which is justified for small
perturbations around a steady state.  When a random fluctuation shifts
the monomer copy number away from its steady-state value, the decay
toward the steady state can be described by the system Jacobian.  The
disparity in the magnitude of the (negative) eigenvalues of the
Jacobian matrix for the MO versus the DA circuits signifies that the
perturbed state is buffered by fast settlement of the monomer-dimer
equilibrium.  This buffering occurs before random fluctuation can
accumulate, possibly with catastrophic physiological effects,
explaining the coarse long-time patterns observed in the MO model in
contrast with the DA circuits (Fig.~2).

\subsection*{Frequency-selective whitening of Brownian noise}
The dimerization process itself generates stochastic fluctuations on a
short time scale.  However, since this time scale is essentially
separated from that of monomer synthesis and decay (orders of
magnitude faster), dimerization effectively mitigates monomer-level
fluctuations.  The frequency content of the fluctuations is best
studied by an analysis of the power spectral density (PSD), which is
defined as the Fourier transform of the autocorrelation
function~\cite{numrec}, originally introduced for signal processing.
Fig.~4 shows the noise power spectra of DA1, and the
distinction between the MO circuit and the DA topology is immediately
evident.  In particular, we note the following two features. (i) A
power-law decay with increasing frequency and (ii) a horizontal
plateau for the DA circuits.  The power-law feature is explained by
the ``random walk'' nature of protein synthesis and decay: The
power-law exponent is approximately 2, which is reminiscent of
Brownian motion (a Wiener process) in the limit of large molecular
copy numbers.  Compared to other commonly observed signals, such as
white (uncorrelated) noise or $1/f$ noise, protein synthesis/decay has
a longer correlation time.  If the autocorrelation function of a time
course is characterized by a single exponential decay, as is the case
for Brownian noise, the PSD is given by a Lorentzian profile, and
thus, well approximated by an inverse-square law in the low-frequency
regime.  We do not observe a saturation value for the MO circuit, and
it is likely not in the frequency window of physiological interest.
This may especially be the case for circuits where the correlation
times are long.

The noise reduction is in the physiologically relevant low-frequency
regime, and in Fig.~4 we have indicated the typical values
for a cell cycle and mRNA lifetime.  Although stochastic fluctuations
impose a fundamental limit in cellular information processing,
multiple noise sources may affect cellular physiology non-additively.
For a living cell, fluctuations are especially relevant when their
correlation time is comparable to, or longer than, the cell cycle.  At
the same time, short-time scale fluctuations (relative to the cell
cycle) are more easily attenuated or do not propagate~\cite{tan07}.
Additionally, the observed flat region in the PSD of the DA circuits
implies that as far as mid-range frequency fluctuations are concerned,
we can safely approximate them as a white noise.  This insight may
shed light on the reliability of approximation schemes for effective
stochastic dynamics in protein-only models.

\subsection*{Increased lifetime of dimer plays an important role}
The virtue of the cytosolic dimer state is also directly related to
the extended lifetime of proteins when in a complex.  Except for the
degradation tagging for active proteolysis, a much slower turnover of
protein oligomers is the norm.  This is partly explained by the common
observation that monomers have largely unfolded structures, which are
prone to be target of proteolysis~\cite{herman03}.  It has also been
pointed out that the prolonged lifetime of the oligomeric form is a
critical factor for enhancing the feasible parameter ranges of gene
circuits~\cite{buchler05}.  As seen from Fig.~3 (also
Table~2), the fold change of the noise reduction, while still
significant, is not as strong for the (hypothetical) case of dimer
lifetime being the same as that of the monomer
($\gamma_2/\gamma_1=1/2$).  However, the low-frequency power spectra
still exhibit almost an order-of-magnitude smaller noise power than in
the MO circuit with the same rate parameters (Fig.~4).
Hence, the noise reduction capability holds good as long as the
dimer lifetime is kept sufficiently long compared with the
monomer-dimer transition.

\subsection*{Effects of homo-dimerization in genetic toggle switch}
The exceptionally stable lysogeny of the phage $\lambda$, for which
the spontaneous loss rate is $\lesssim10^{-7}$ per cell per
generation~\cite{rozanov98,little99}, has motivated the synthesis of a
genetic toggle switch~\cite{gardner00}. Toggle switch is constructed 
from a pair of genes, which we will denote as gene $A$ and $B$, that
transcriptionally repress each other's expression.  This mutual
negative regulation can be considered an effective positive feedback
loop and provides the basis for the multiple steady states.  
The existence of multistability, in turn, may be exploited as a device 
for epigenetic memory or for decision making~\cite{losick08}. 
 
As the general attributes of positive feedback with cooperativity
suggest, a genetic toggle switch responds to external cues in an
ultrasensitive way: When the strength of a signal approaches a
threshold value, the gene expression state can be flipped by a small
change in the signal.  For example, the concentration of protein $A$
($B$) may rapidly switch from high to low and vice versa.  However,
previous studies of a synthetic toggle switch have shown that the
noise-induced state switching is a rare
event~\cite{kobayashi04,tian06,gardner00}.  In the ensuing analysis,
we aim to delineate the origin of this exceptional stability.

In a simple model, the monomer-only (MO) toggle, regulatory proteins
only exist in monomeric form.  Although an external signal is not
explicitly included, random fluctuations in the abundance of the
circuit's molecular components will occasionally flip the toggle-state
for the two protein species.  Drawing on the results from our analysis
of positive autoregulatory gene circuits, we hypothesize that
dimerization in the regulatory proteins of the toggle switch will
serve to stabilize its performance against noise.  We allow the
protein products of each gene to form a homodimer, being either $AA$
or $BB$, which is similar to the cI-cro system in phage
$\lambda$~\cite{ptashne}.  The dissociation constant for the dimers is
defined as $K_1=q_1/k_1$, where $k_1$ is the rate of two monomers
forming a complex, and $q_1$ the rate of the complex breaking up into
its two constituents.  

We evaluate the effect of the fast protein binding-unbinding dynamics
on the toggle switch performance by using either (i) the monomers or
(ii) the homodimers as the functional form of the repressor.
Fig.~5 shows, for selected values of the dissociation
constant $K_1$, representative time series of the protein monomer
(left) and dimer (right) abundances for the case of (a) monomeric or
(b) dimeric transcription factors, respectively.  A careful analysis
of the phase space (in presence of noise) for our chosen set of
parameters confirms that the studied toggle-switch systems are in the
bistable region~\cite{ghim08}.

When monomer is the functional form of the repressor molecule 
(Fig.~5(a)) and $K_1$ is large (limit of low
dimer affinity), the protein populations are dominated by monomers.
Hence, the circuit effectively behaves as an MO toggle.  As $K_1$
decreases, we see that the level of random switching is suppressed:
Analogous to the autogenous circuit, the dimer pool stabilizes the
protein monomer population.  However, the noise suppression is not
monotonic with increasing dimer binding affinity.  Indeed, for very
large binding affinities (small $K_1$), the number of random switching
events is increased since the monomer is only available in low copy
numbers.  Consequently in this limit, it becomes more likely that a
small fluctuation in the monomer abundance can cause a dramatic change
in the overall gene expression profile.  The noise-stabilizing effect
of dimerization is also reflected in the corresponding PSDs 
[Additional file 1].  For instance, we observe a
marked suppression of low-frequency fluctuations in the monomer
abundance with increasing $K_1$.

In Fig.~5(b) we show corresponding sample time series
for the case of a dimeric repressor, all other properties being the
same as in (a).  While the overall trends are similar, we do note the
following difference.  Contrary to the monomeric repressor case, there
are very few toggle events in the strong binding limit: Since the
signaling molecules (dimers) of the dominant gene (the ``on''-gene)
tend to exist in large copy numbers, a significant fluctuation is
needed to flip the state of the toggle switch.  In the case of
monomeric repression, the signaling molecule exists in low abundance
in this limit.  Thus, the dominant protein species in the
dimeric-repressor system is able to maintain much better control over
the state of the toggle switch.

In Fig.~6, we show the distribution $(N_A-N_B)$,
the difference in molecule abundance for the two protein species in
the case of monomeric (left) and dimeric (right) transcription factor.
The asymmetry with respect to the zero axis is caused by our choice of
initial conditions (protein species $A$ in high concentration and
species $B$ in low concentration), as well as the finite length of the
time series.  For monomeric transcription, the presence of dimers with
moderate binding affinity sharpens the monomer abundance distribution
while accentuating its bimodal character.  This is in agreement with
the qualitative observation from Fig.~5 on switching
stability.  For dimeric transcription, we clearly observe that the
symmetry of the system is broken for small values of $K_1$, indicating
that the state of the toggle switch is extremely stable, and hence,
likely determined by the choice of the initial conditions.

To systematically quantify our observations on the interplay between
dimer-binding affinity and the functional stability of the toggle
switch, we generated long time series ($\approx3\cdot10^7$ sec) to
measure the average spontaneous switching rate.  In
Fig.~7, we show the average toggle frequency relative
to that of the MO toggle for the binding affinities 
$K_1/\textrm{nM}=\{2, 20, 100,1000\}$, and the average MO switching 
rate is $7.5\times10^{-6}$/hour. As expected, we find that intermediate 
values of $K_1$ are able to stabilize the toggle switch.  
Fig.~7 also highlights the increased stability of the 
toggle switch for a dimeric versus monomeric transcription factor, 
the dimeric switching rates always being lower and approaching zero 
for strong dimer binding.

\subsection*{Heterodimerization in genetic toggle switch}
We have also considered the case of heterodimerization in the toggle
switch, since the noise- and functional stabilization of the switch
may be directly affected by the composition and source of the dimers.
Note that, the gene-regulation activity is conferred by the two
monomer proteins $A$ and $B$ and not the heterodimer $AB$.  However,
we find that the presence of (inactive) heterodimers gives rise to
very similar noise-stabilizing effects as that of homodimers
(Fig.~7).  In fact, the existence of heterodimer state 
allows the dominant protein species to effectively suppress the (active) 
monomers of the minority species. Thus the heterodimer circuit shows
dramatically enhanced functional stability as compared to the case of
homodimeric repressors, not sharing the discussed vulnerability of MO 
circuit to intrinsic noise. 
Although, to our knowledge, this is a purely hypothetical toggle-switch 
design, it provides a general strategy for noise control in synthetic gene
circuits, along with previously proposed approach of overlapping
upstream regulatory domains~\cite{warren04}.

\section*{Conclusions}
Cells have evolved distinct strategies to combat the fundamental
limits imposed by intrinsic and environmental fluctuations.  We
investigated the role of protein oligomerization on noise originating
from the random occurrence of reaction events and the discrete nature
of molecules.  Recent efforts to correlate network structure with
functional aspects may provide valuable insights into approaches for
network-level noise control~\cite{barabasi04}.  While negative
feedback is one of the most abundantly observed patterns to achieve
the goal of stability, it begs the question of how cells reliably
change the expression of genes from one state to another.  The
ultrasensitive response circuit, exemplified by the ubiquitous signal
transduction cascades in eukaryotic cells, has been proposed as an
answer to this question~\cite{goldbeter81,huang96}.

In addition to the combinatorial expansion of functional specificity,
we argue that the availability of oligomeric states contributes to the
attenuation of stochastic fluctuations in protein abundance.  In
positive autoregulatory gene circuits, where the abundance of an
expressed protein controls its own synthesis rate, dimerization
provides a buffer serving to mitigate random fluctuations associated
with the bursty transcription-translation process. We find that
short-time binding-unbinding dynamics reduce the overall noise level
by converting potentially pathological low-frequency noise to
physiologically unimportant, and easily attenuated, high-frequency
noise~\cite{tan07}.

Noise-induced switching generally signals a defect in cellular
information processing.  Untimely exit from latency in the
lambda-phage system directly implies, as the immediate consequence to
viruses, increased chance of being targeted by a host immune system.
In the case of a bacterium, the expression of a specific set of sugar
uptake genes when the sugar is absent from the external medium is a
considerable waste of cellular resources.  For example, \textit{lac}
operon of \textit{E. coli} can be considered to have the circuitry of
mutual antagonism between the \text{lacI} gene and lactose
uptake-catabolic genes~\cite{lacoperon}. A difference lies in the 
non-transcriptional deactivation of the allosteric transcription 
factor LacI.  LacY, lactose permease, indirectly regulates LacI 
by increasing lactose uptake, which in turn catalytically deactivates LacI.  
Likewise, many pili operons of Gram-negative bacteria are also known 
to utilize heritable expression states, which are of crucial role in
pathogenesis~\cite{hernday02,sauer00}.

We expect that the random flipping of gene expression states in the
examples of positive-feedback-based genetic switches may very well be
closely coupled with the fitness of an organism.  Phenomenological
models relating the fitness of an organism to random phenotypic
switching in fluctuating environments have provided important insights
into the role of noise~\cite{thattai04}, but still many questions
remain unanswered.

Applying these insights to the design of a synthetic gene switch
demonstrates the potential use of affinity-manipulation for synthetic
biology, where the construction of genetic circuits with tunable
noise-resistance is of central importance.  In particular, our
analysis highlights the potential utility of heterodimerization to
stabilize ultrasensitive switches against random fluctuations.  In
practice, small ligand molecules may be employed to regulate and tune
the binding affinity of regulatory proteins, being either monomers or
dimers.  Our results further suggest that the structure of the
protein-interaction network~\cite{losick08} may provide important
insights on methods for genome-level noise control in synthetic and
natural systems.

%%%%%%%%%%%%%%%%%%
\section*{Methods} 
\subsection*{Model construction}
To evaluate the general role of protein oligomerization in a broad
functional context, we studied the two most common motifs found in
genetic regulatory circuits: positive autoregulation and the bistable
switch.  The reaction scheme studied is
summarized in Fig.~1, where the binding/unbinding
reactions between RNAp and promoter or between TF and operator are
made explicit. 
Each distinct binding status of DNA is associated with 
a unique transcription initiation rate, and then the overall rate of 
mRNA synthesis is a weighted average of the initiation rates for 
distinct binding status, where the weights are given by the relative 
abundance of each configuration at equilibrium, determined by the 
calculation of binding energy~\cite{gilman02}. 
Note that, neither binding equilibrium nor empirical
Hill-type cooperativity is assumed \textit{ad hoc}.  In particular, we
split the lumped transcription process into two separate events, (i)
isomerization of closed RNAp-promoter complex to its open form and
(ii) transcription elongation followed by termination.  This is to
reflect the availability of the free promoter while the transcription
machinery proceeds along the coding sequence of a gene as soon as the
promoter region is cleared of the RNAp holoenzyme.  Otherwise, the
promoter would be inaccessible during a whole transcription event,
altering the random mRNA synthesis dynamics.

To realize the genetic toggle switch in a stochastic setting, we keep
track of the microscopic origin of cooperativity that gives rise to
bistability.  Among various strategies, we employ multiple operator
sites which have the same binding affinity with the repressor.  The
resultant circuitry is, in essence, two autogenous circuits, $A$ and
$B$, which are connected through the active form of their expressed
proteins (the active form being either monomer or dimer).  The
connection is implemented by allowing the active form of proteins $A$
($B$) to bind the operator sites of gene $B$ ($A$).  In order to make
the interaction between the two genes repressive, unlike positive
autogenous circuits, $K_{31}$ and $K_{32}$ in Fig.~1 are
now greater than $K_{30}$, making the protein transcriptional
repressors.  For reasons of analytical simplicity, we have studied the
symmetric toggle switch, where the reaction descriptions of each
component follow those of the autogenous circuit.  Again, the
quantitative characteristics of macromolecular binding-unbinding are
chosen based on the phage lambda-\textit{E. coli} system. The only
exception is related to the multiple operator sites, where the second
repressor binds an operator site with higher binding affinity when the
first site is already occupied by the repressor protein~\cite{tian04}.
We introduce three different dimerization schemes.  Three different
dimerization scheme have been introduced: (i) homodimerization with
monomeric repressor, (ii) homodimerization with dimeric repressor, and
(iii) heterodimerization with monomeric repressor.  By solving for the
stationary states of the deterministic rate equations, we could
identify the bistability region in parameter space to which all the
model systems under consideration belong.  

\subsection*{Stochastic simulation}
While the deterministic rate equation approach or Langevin dynamics 
explicitly gives the time-evolution of molecular concentration in the 
form of ordinary differential equations, chemical master equation 
(CME) describes the evolution of a molecular number state as a 
continuous-time jump Markov process.
To generate the statistically correct trajectories dictated by CME,
we used the Gillespie direct~\cite{gillespie77} and Next Reaction
(Gibson-Bruck)~\cite{gibson00} algorithms, both based on the exact
chemical master equation.  The Dizzy package~\cite{dizzy} were used as
the core engine of the simulations.  To ensure that calculations were
undertaken in a steady state, we solved the deterministic set of
equations for steady state using every combination of parameters
investigated.  We employed these deterministic steady-state solutions
as initial conditions for the stochastic simulations.  For each model
system, we generated $10^5$ ensemble runs with identical initial
conditions and used the instantaneous protein copy number at a fixed
time point $t=5000$ sec.  To achieve high-quality power spectra in the
low- and high-frequency limits, we ran time courses ($\sim 10^5$ sec)
with higher sampling frequency (20 measure points per sec).

To calculate the average switching rate, we generated time series of
minimum length $3\cdot10^7$ sec (approximately corresponding to 1
year).  We identify a state change in the toggle switch by monitoring
the ratio of the monomer and dimer abundance for the two protein
species.  In order to avoid counting short-time fluctuations that do
not correspond to a prolonged change of the toggle state, we a applied
sliding-window average to the time series, using a window size of
$1000$ sec.

%%%%%%%%%%%%%%%%%%%%%%%%%%%%%%%%
\section*{Authors' contributions}
CMG and EA designed the study. CMG performed the computations.  CMG
and EA analyzed the results and wrote the paper.  Both authors have
read and approved the final version of the paper.
   
%%%%%%%%%%%%%%%%%%%%%%%%%%%
\section*{Acknowledgments}
\ifthenelse{\boolean{publ}}{\small}{} The authors thank Dr. Navid for
thoughtful discussion and suggestions.  This work was performed under
the auspices of the U. S. Department of Energy by Lawrence Livermore
National Laboratory under Contract DE-AC52-07NA27344 and funded by the
Laboratory Directed Research and Development Program (project
06-ERD-061) at LLNL.

%% References 
%%%%%%%%%%%
%{\ifthenelse{\boolean{publ}}{\footnotesize}{\small}

\bibliographystyle{bmc_article}  % Style BST file
\bibliography{dimer}             % Bibliography file (usually '*.bib' ) 
\ifthenelse{\boolean{publ}}{\end{multicols}}{}

%% Tables
%%%%%%%%%%%%%%%%%%%%%%%%%%%%%%%%%%%
%% Use of \listoftables is discouraged.
\newpage
\noindent
{\sffamily\LARGE Tables}

\vskip 1cm
\noindent
\textbf{Table 1 - Probability rates for positive autogenous circuit.}\\
\vskip -0.2cm
\begin{table}[h!]
\begin{center}
\begin{tabular}{cclll}
\toprule
\hline
Category & Symbol  & ~~~~~Reaction    & ~~Value (s$^{-1}$) & ~~~Ref.    \\ 
\hline
\multirow{2}{*}{protein dimerization}&
$k_1$       & P$_1$ $+$ P$_1 \rightarrow$ P$_2$ & 0.001-0.1 &\multirow{2}{*}{\cite{sauer79,arkin98}} \\
&$q_1$      & P$_2 \rightarrow$ P$_1$ $+$ P$_1$ & 0.1-1   & \\
\hline
\multirow{6}{*}{TF-operator int}&
$k_{20}$    & P$_2$ $+$ D00 $\rightarrow$ D20 & 0.012 &\multirow{6}{*}{\cite{ackers82,hawley82,hawley85}}\\
&$q_{20}$   & D20 $\rightarrow$ P$_2$ $+$ D00 & 0.9   & \\
&$k_{21}$   & P$_1$ $+$ D00 $\rightarrow$ D10 & 0.038 & \\
&$q_{21}$   & D10 $\rightarrow$ P$_1$ $+$ D00 & 0.3   & \\
&$k_{22}$   & P$_1$ $+$ D10 $\rightarrow$ D20 & 0.011 & \\
&$q_{22}$   & D20 $\rightarrow$ P$_1$ $+$ D10 & 0.9   & \\
\hline
\multirow{6}{*}{RNAp-promoter int}&
$k_{30}$    & R $+$ D00 $\rightarrow$ D01     & 0.038 &\multirow{6}{*}{\cite{dehaseth98,ujvari96,shea85}}\\
&$q_{30}$   & D01 $\rightarrow$ R $+$ D00     & 0.3  & \\
&$k_{31}$   & R $+$ D10 $\rightarrow$ D11     & 0.038$^\dagger$, 0.38$^\ddagger$ & \\
&$q_{31}$   & D11 $\rightarrow$ R $+$ D10     & 0.3$^\dagger$, 0.03$^\ddagger$ & \\
&$k_{32}$   & R $+$ D20 $\rightarrow$ D21     & 0.38$^{*\dagger}$ & \\
&$q_{32}$   & D21 $\rightarrow$ R $+$ D20     & 0.03$^{*\dagger}$ & \\
\hline
Isomerization &$v$
            & Dx1 $\rightarrow$ C $+$ Dx0     & 0.0078 &\multirow{1}{*}{\cite{hawley85}}\\
\hline
\multirow{5}{*}{tsx-tsl elongation \& decay}&
$\alpha$    & C $\rightarrow$ M $+$ R         & 0.03   &\multirow{5}{*}{\cite{alberts02,lewin04}} \\
&$\beta $   & M $\rightarrow$ P$_1$ $+$ M     & 0.044  & \\
&$\gamma_0$ & M $\rightarrow \varnothing$     & 0.0039 & \\
&$\gamma_1$ & P$_1 \rightarrow \varnothing$   & 7$\times$10$^{-4}$ & \\
&$\gamma_2$ & P$_2 \rightarrow \varnothing$   & 0.7-3.5$\times$10$^{-4}$ & \\ 
\hline
\bottomrule
\end{tabular}
\label{rates}
\end{center}
\end{table}
\vskip -0.5cm
\noindent
Kinetic rates for the positive autogenous circuit. Experimentally
available rates are all taken from lambda phage-\textit{E. coli} complex. 
The values with superscript correspond to the circuit topologies DA1 (*), DA2
($\dagger$), and DA3 ($\ddagger$) in Fig.~1.

\vskip 2cm

\noindent
\textbf{Table 2 - Relative Fano factors of protein abundance distributions}\\
\vskip -0.2cm
\begin{table}[h!]
\begin{center}
\begin{tabular}{ccccc}
\toprule
\hline
\multirow{2}{*}{$K_1$ (nM)}&
\multicolumn{2}{c}{$\gamma_2=\gamma_1/10$}&\multicolumn{2}{c}{$\gamma_2=\gamma_1/2$}\\
\cline{2-5} 
&monomer&dimer&monomer&dimer\\
\hline
1   & 0.127 & 0.809 & 0.132 & 0.679\\
20  & 0.209 & 0.936 & 0.230 & 0.716\\
500 & 0.866 & 0.478 & 0.826 & 0.426\\
\hline
\bottomrule
\end{tabular}
\label{fano}
\end{center}
\end{table}
\vskip -0.5cm
\noindent
The Fano factor of protein abundance distribution for the autogenous circuits 
(topology DA1), relative to that of the monomer-only (MO) circuit, $8.729$.

\newpage
\noindent
{\sffamily\LARGE Figures}

\vskip 1cm

\noindent
\textbf{Figure 1. Schematic of model autoregulation gene circuit.}\\
\noindent
The DNA binding status is indicated by {\sffamily Dxy}, where {\sffamily x}
corresponds to the operator region (empty=0, monomer=1, dimer=2), and
{\sffamily y} to the promoter region (empty=0, RNA polymerase
bound=1).  {\sffamily C} represents the open complex of DNA-RNAp
holoenzyme with the promoter sequence just cleared of RNAp and is
subject to transcription elongation.  Finally, {\sffamily M},
{\sffamily P1} and {\sffamily P2} correspond to mRNA, protein monomer,
and dimer, respectively.  The network topologies can be grouped into
two classes, monomer-only (MO) or dimer-allowed (DA) circuits. We have
studied DA1 (red lines), which only allows the dimer to bind with the
DNA-operator sequence, DA2 (green) with sequential binding of monomers
on the DNA, and DA3 (blue), which shares protein-DNA binding kinetics
with MO while allowing dimerization in the cytosol. Note that for
topology DA2, we have chosen $K_{31}=K_{30}$ (see text for details) We
have assumed cells to be in the exponential growth phase and the
number of RNAp ({\sffamily R}) constant.

\vskip 1cm

\noindent
\textbf{Figure 2. Ten independent time courses of the abundance of protein
monomers in the (positive) autoregulatory circuit.}\\
\noindent
The availability of a cytosolic dimer state (red, using circuit topology DA1)
significantly reduces the copy-number fluctuations of the monomer
compared to the monomer-only (MO) circuit (blue).  All corresponding
MO and DA1 parameters have the same values. In the ensuing simulations
initial conditions are chosen to be the steady state solution of the 
corresponding deterministic rate equation so that the transient 
behavior should be minimized.

\vskip 1cm

\noindent
\textbf{Figure 3. Stationary state distribution of monomer (black) and dimer
(orange) protein abundance in the positive autogenous circuits.} \\ 
\noindent
The left (right) column corresponds to a ratio of the dimer and monomer
decay rates of $\gamma_2/\gamma_1=1/10$ ($\gamma_2/\gamma_1=1/2$).
The molecular copy numbers are collected at a fixed time interval
($5\cdot10^3$ sec) after the steady state has been reached.  Here
$K_1\equiv q_1/k_1$ is the dissociation constant of the protein dimer.
As the binding equilibrium is shifted towards the dimer state
(decreasing $K_1$), the noise level is monotonically reduced (see
Table 2).  Note that the prolonged protein lifetime due to the complex
formation (left column) affects the noise level.

\vskip 1cm

\noindent
\textbf{Figure 4. Power spectral density (PSD) of fluctuations in protein
abundance.}\\  
\noindent
The PSD of the MO circuit clearly displays a power-law
behavior.  All other model systems with an available cytosolic protein
dimer state (DA1 shown here) develop a plateau in the mid-frequency
region regardless of the model details (see \textit{Supplementary
Information}).  As the dimer binding affinity increases, the noise
level is further reduced. We have included the MO result in the dimer
panel (right) for reference.  Datasets with solid (empty) symbols
correspond to $\gamma_2/\gamma_1=1/10$ ($\gamma_2/\gamma_1=1/2$).

\vskip 1cm

\noindent
\textbf{Figure 5. Sample time series of monomer and dimer copy numbers 
in genetic toggle switch.}\\
\noindent
(a) MTF circuit, where monomer is the functional form of the repressor.
(b) DTF circuit, where dimer is the functional form of the repressor.
The left (right) column shows the number of the two monomer molecules 
$A$ and $B$ (dimers $AA$ and $BB$), and the initial state is always with 
species $A$ (red) in high abundance.  Note that the switching frequency 
depends on the binding affinity of protein dimer.

\vskip 1cm

\noindent
\textbf{Figure 6. Distribution of monomer abundance differences 
between protein species $A$ and $B$.}\\
\noindent
The asymmetry with respect to
the zero axis is due to the choice of initial state (species $A$ high)
and the finite time span of simulations.

\vskip 1cm

\noindent
\textbf{Figure 7. Random switching rates of genetic toggle switches.}\\  
\noindent
Ordinate is the ratio of the random switching rates of various toggle 
switches to that of the monomer-only (MO) circuit, $7.5\times 10^{-6}$/hour.
{\sffamily MTF}, monomeric transcription factor; 
{\sffamily DTF}, dimeric transcription factor;
{\sffamily Het-MTF}, monomeric transcription factor with deactivated 
heterodimer state. 

\newpage
\noindent
{\sffamily\LARGE Additional Files}

\vskip 1.5cm

\noindent
\textbf{Additional file 1. Supplementary results for the positive 
autoregulatory circuits with various topology.}\\
\noindent
Protein abundance distribution and power spectral density of autogenous 
DA2 and DA3 circuits are presented. 
\end{bmcformat}
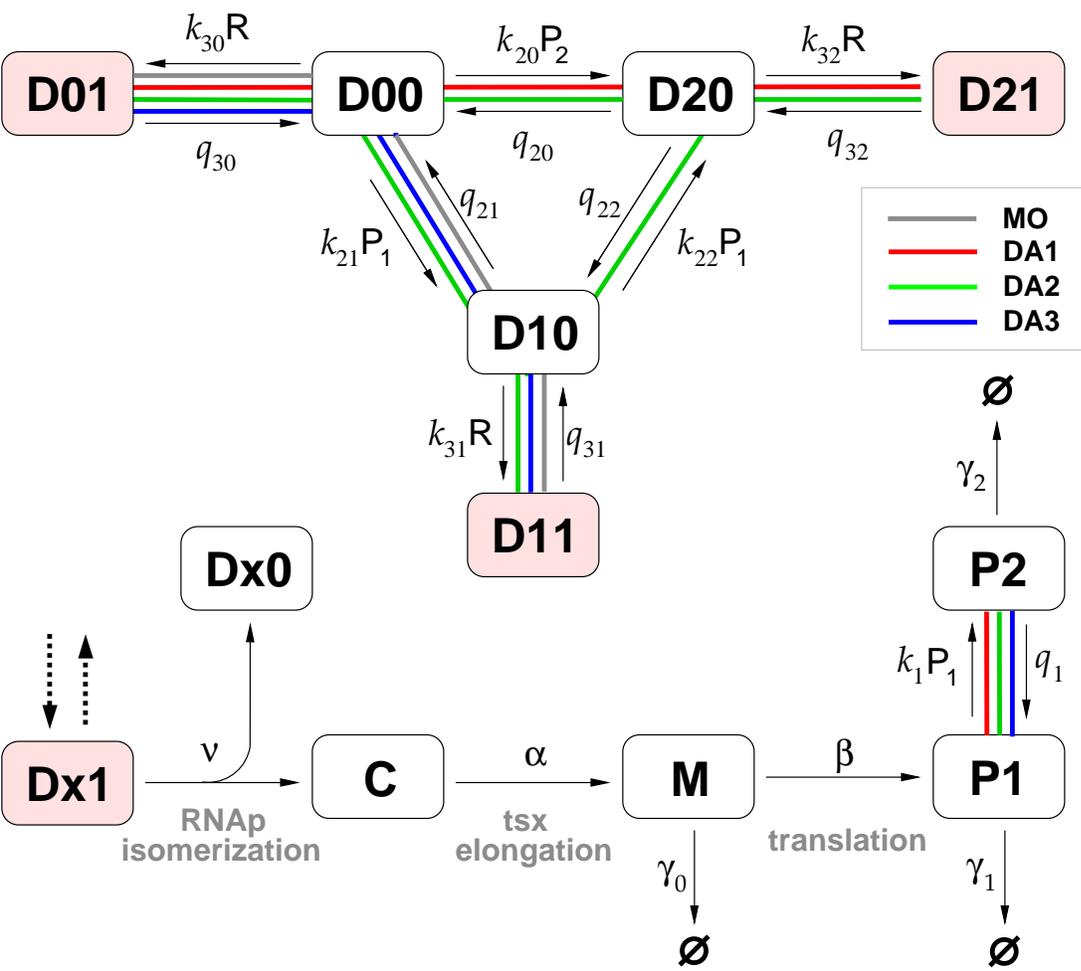
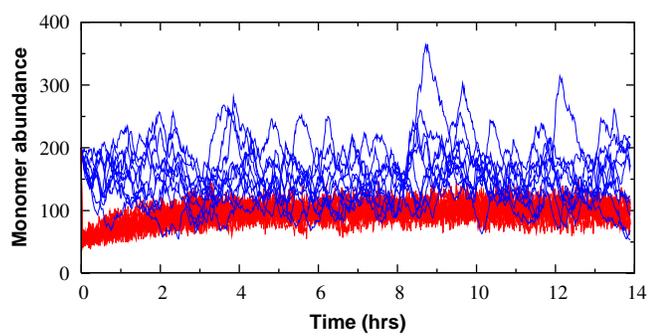
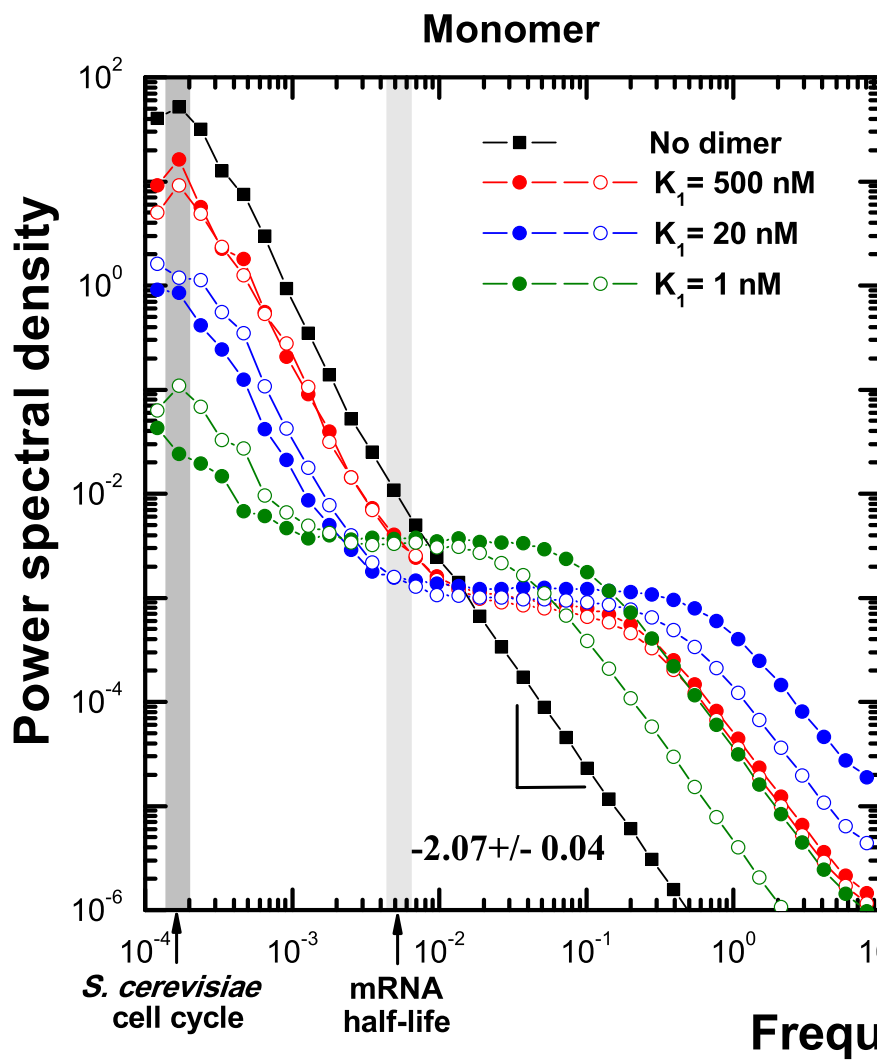
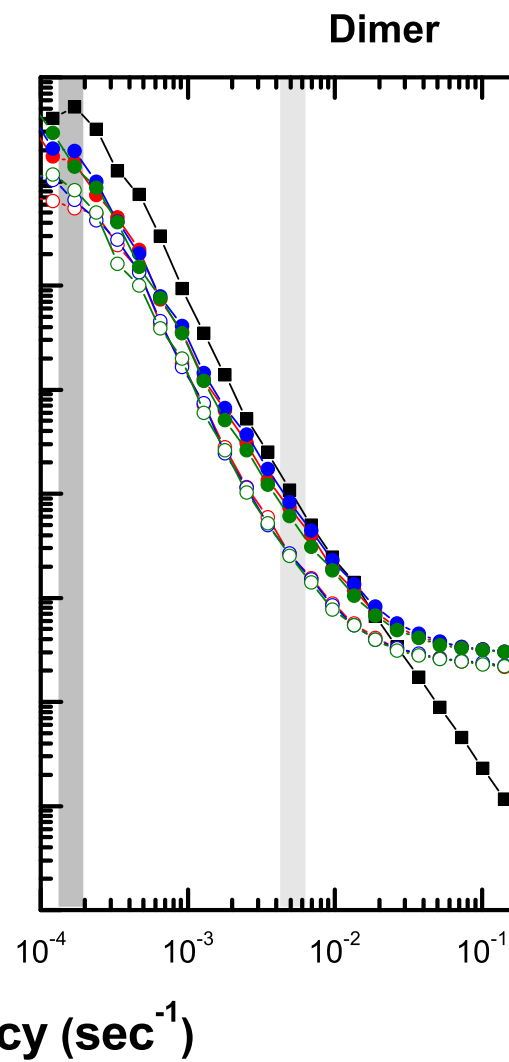
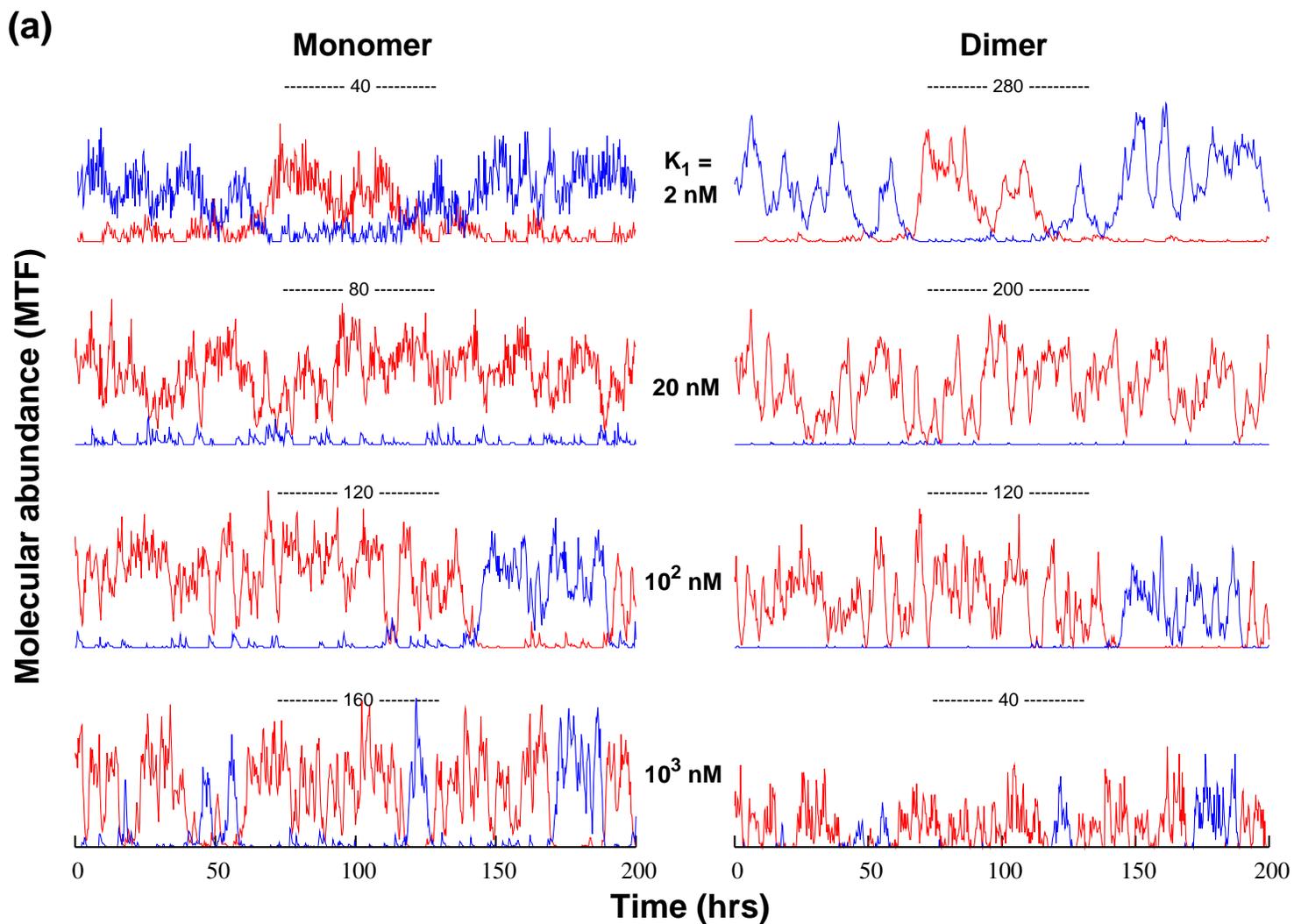
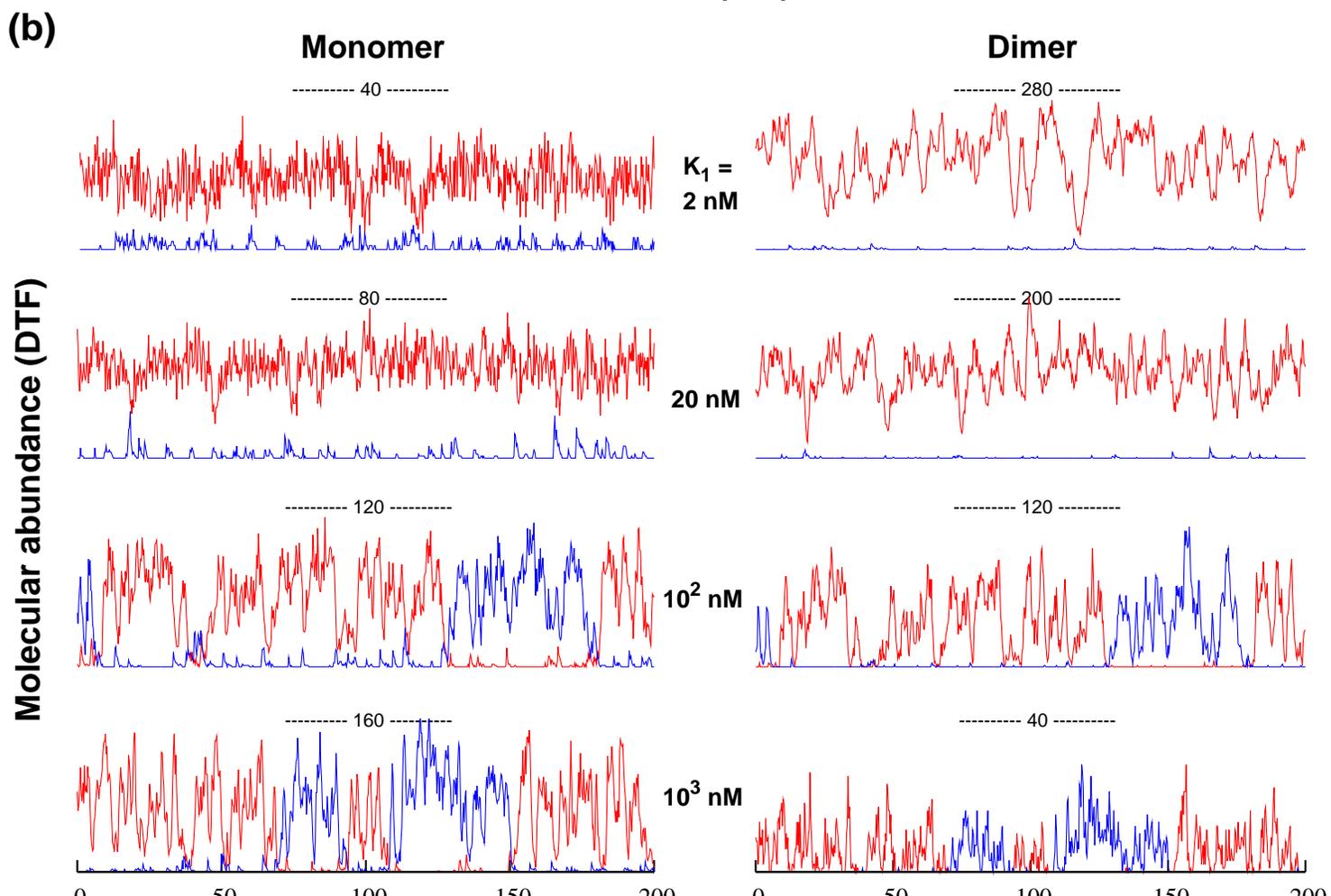
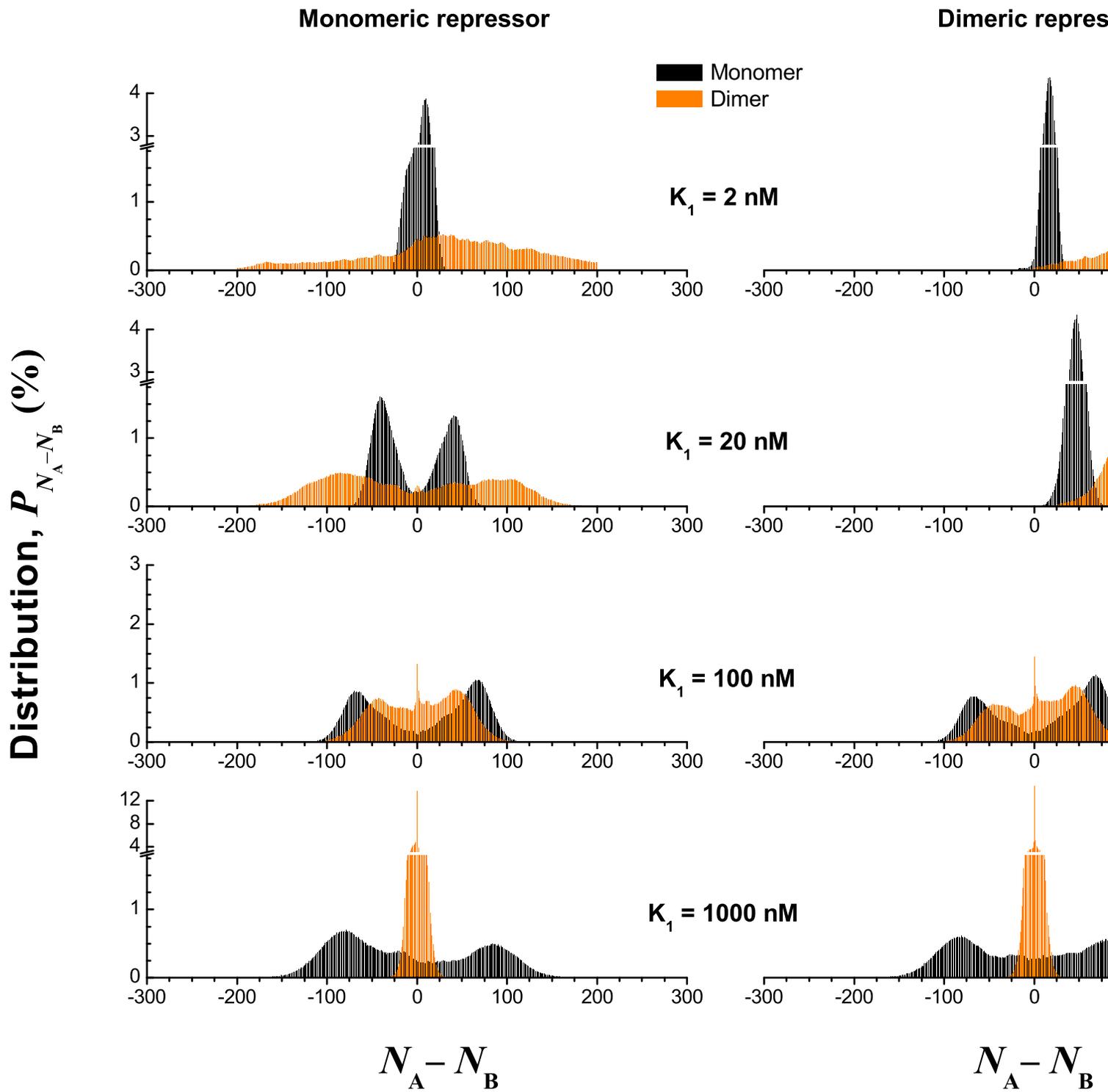
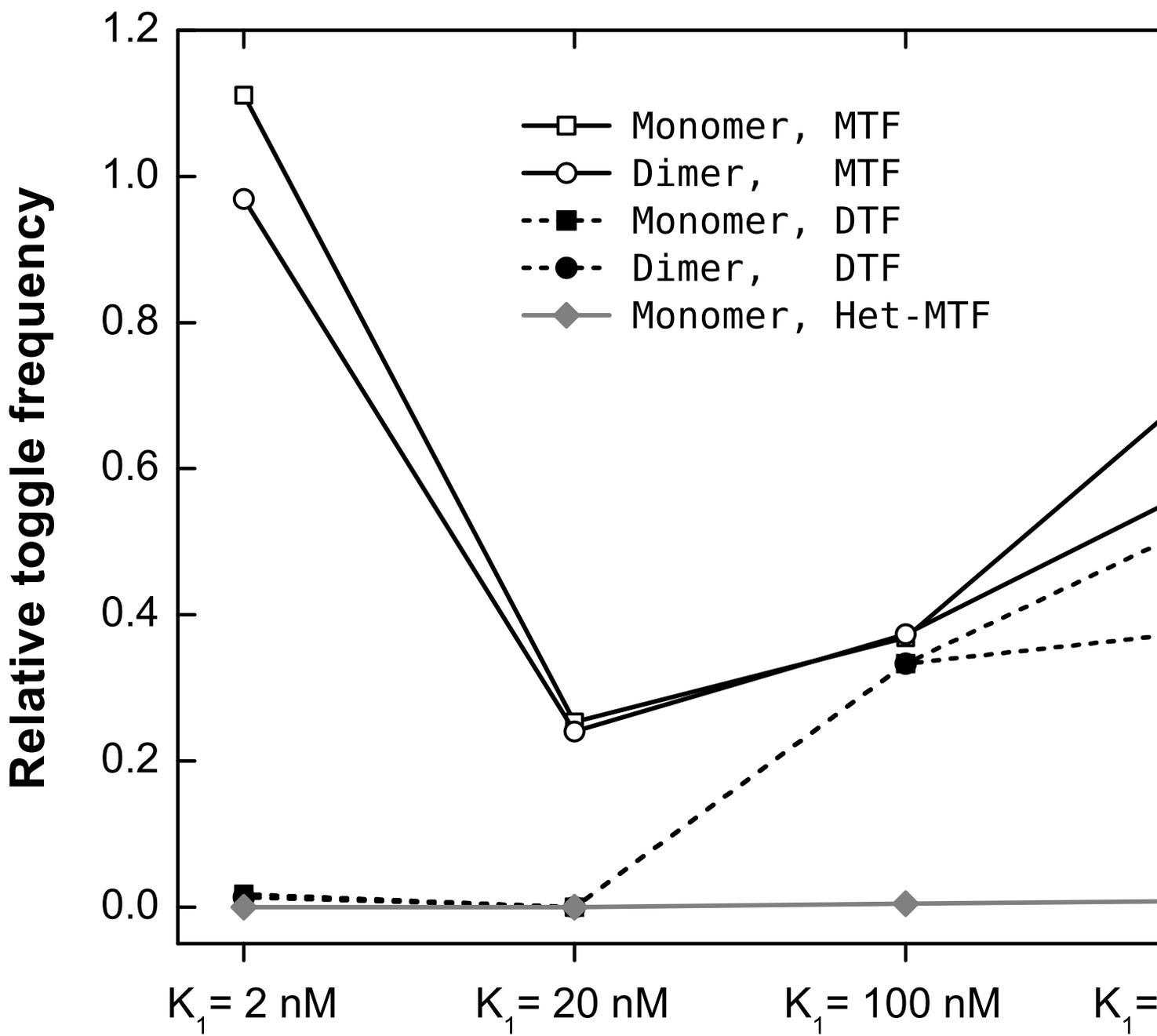
\end{document}